\documentclass{jetpl}
\usepackage{epsfig,cite}
 \twocolumn
\lat

\title{ Superconducting spin filter}
\rtitle{Superconducting spin filter} \sodtitle{Superconducting
spin filter}

\author{N.\,M.~Chtchelkatchev$^{\sharp\S}$\/\thanks{e-mail: nms@landau.ac.ru},}
\rauthor{N.\,M.~Chtchelkatchev }

\sodauthor{N.\,M.~Chtchelkatchev}

\address{$^{\sharp}$L.\,D.~Landau Institute for Theoretical Physics RAS,
117940 Moscow, Russia\\
$^{\S}$ Institute for High Pressure Physics, Russian Academy of
Sciences, Troitsk 142092, Moscow Region, Russia }
\dates{June 22, 2003}

\abstract{Consider two normal leads coupled to a superconductor;
the first lead is biased while the second one and the
superconductor are grounded. In general, a finite current
$I_2(V_1,0)$ is induced in the grounded lead 2; its magnitude
depends on the competition between processes of Andreev and normal
quasiparticle transmission from the lead 1 to the lead 2. It is
known that in the tunneling limit, when normal leads are weakly
coupled to the superconductor, $I_2(V_1,0)=0$, if $|V_1|<\Delta$
and the system is in the clean limit. In other words, Andreev and
normal tunneling processes compensate each-other.  We consider the
general case: the voltages are below the gap, the system is either
dirty or clean. It is shown that  $I_2(V_1,0)=0$ for general
configuration of the normal leads; if the first lead injects spin
polarized current then $I_2=0$, but spin current in the lead-2 is
finite. XISIN structure, where X is a source of the spin polarized
current could be applied as a filter separating spin current from
charge current. We do an analytical progress calculating
$I_1(V_1,V_2),\,I_2(V_1,V_2)$.}


\begin{document}
\maketitle

\begin{figure}[tb]
\begin{center}
\includegraphics[width=80mm]{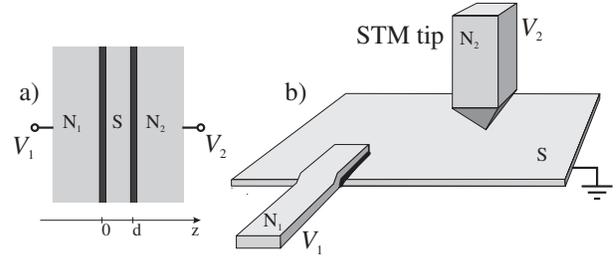}
\end{center}
\caption{Fig1. The outline of the setup. N$_{1,2}$ are normal
metals or ferromagnets.} \label{fig1}
\end{figure}

Hybrid systems consisting of a superconductor (S) and two or more
normal metal (N) or ferromagnetic (F) probes recently started to
attract a great attention
\cite{Deutscher,Falci,Melin,Lesovik,Jedema}. Among most striking
new results is the prediction that NSN (FSF) devices can play the
role of entangler  producing Einstein-Podosky-Rosen (EPR) pairs
\cite{Lesovik} having potential applications, for example, in
quantum cryptography \cite{q_information}. Not long ago rather
unusual effect was described in normal metal  -- tunnel barrier
(I) -- superconductor -- tunnel barrier -- normal metal (NISIN)
junction (see, e.g., Fig.1b) \cite{Falci,Melin}. It was shown that
when N$_1$ is biased, N$_2$ and S are grounded there is no current
injection from N$_1$ to N$_2$ at subgap biases; main assumptions
were: 1) the superconductor is clean, 2) large number of
conducting channels are involved in electron tunneling through NS
interfaces \cite{Falci,Melin}. In other words, the subgap cross
conductance $G_{12}\equiv\partial_{V_1}
I_2(V_1,0)|_{|V_1|<\Delta}=0$, where the current $I_1$ flows in
N$_1$, $V_1$ is the bias between N$_1$ and S and $V_2$
--- between N$_2$ and S. The suppression of $G_{12}$ was attributed
to the compensation of the contributions to the current from
Andreev and normal quasiparticle tunneling processes between N$_1$
and N$_2$ \cite{Falci}. It was also noted that $G_{12}\neq 0$ in
FISIF junctions: $G_{12}$ decays exponentially as $\exp(-r/\xi)$
with the characteristic distance $r$ between the normal terminals
(see, e.g., Fig.\ref{fig1}b), where $\xi$ is the superconductor
coherence length; at small $r/\xi$, $G_{12}$ decays also rather
quickly (at atom-scales): as $1/(k_F r)^2$ ($k_F$ in the
superconductor) \cite{Falci}. Thus with clean superconductors a
measurement of $G_{12}$ may become difficult.

In this letter we first of all  generalize results
\cite{Falci,Melin} and get rid of the assumption 1) [i.e., S is
not restricted to be clean]. We show that when the superconductor
is dirty (the mean free path is smaller than $\xi$) Andreev and
normal transmission rates [as well as $G_{12}$ in FISIF junctions]
\textit{slowly} decay with the characteristics distance $r$
between the normal (ferromagnetic) terminals (at $r<\xi$) in
contrast to the clean regime (see Refs. \cite{Falci,Melin}). For
example, in FISIF with superconducting layer thinner than $\xi$,
see Fig.~\ref{fig1}b, $G_{12}\sim \ln(r/\xi)$; when the
superconductor is bulk then $G_{12}\sim \xi/r$ [$r>\lambda_F$ is
supposed]. Measurements of the effects, related to electron
tunneling through a superconductor (e.g., $G_{12}$) in the dirty
superconductor case can be easier realized experimentally then in
the clean case because then $r$ is not restricted to atomic scales
but by $\xi\gg\lambda_F$.  We show that contributions to the
current from Andreev and normal quasiparticle tunneling processes
always compensate each other in NISIN junctions (so, e.g.,
$I_2(V_1,V_2=0)=0$ for $|V_1|<\Delta$ in first nonvanishing order
over the transparencies of the layers I) for any amount of
disorder in the S-layer. If one prepares a NISIN junction with
layers I having large transparency then normal tunneling start
dominating Andreev tunneling (and $I_2(V_1,V_2=0)\neq 0$). We also
considered FISIN junction, in particular with $V_F\neq 0$ and
$V_N=0$. Then the ferromagnet F plays the role of the
spin-polarized current injector. In this case $I_2(V_F,V_N=0)=0$
also, but spin current in N is finite: charge component of the
current converts into the supercurrent, spin accumulates in N. So
XISIN structure, where X is a source of the spin polarized
current, could be applied in spintronics \cite{Fazio} as a filter
separating spin current from charge current. We find Andreev
$T_{he}$ and normal transmission probabilities $T_{ee}$ of a NISIN
sandwich for subgap energies $|E|<\Delta$ and different angles
$\theta$ between incident quasiparticle trajectory and the normal
to NS interface. It is shown that  the probabilities have
resonances where $T_{he}\sim T_{ee}$; averages of $T_{he}$ and
$T_{ee}$ over incident channels (over $\theta$) are equal  ---
this is the reason why $I_2(V_1,V_2=0)$ is suppressed and the spin
current $I_2^{(s)}(V_1,V_2=0)$ is finite.

We start investigation of NISIN structures from the sandwich
sketched in Fig.\ref{fig1}a: barriers at NS boundaries provide
spectacular reflection; electrons in N and S move ballistically;
the number of channels at both NS boundaries is much larger than
unity.
\begin{figure}[tb]
\begin{center}
\includegraphics[width=70mm]{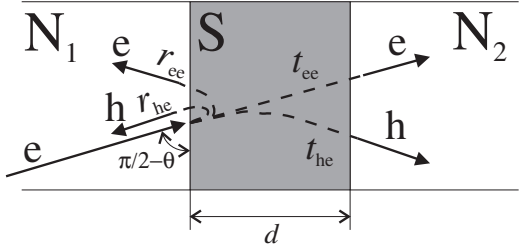}
\caption{\label{fig2} Fig2. Electron scattering from a NSN
junction.}
\end{center}
\end{figure}
The transmission probabilities $T_{he}(E,\theta)$ and
$T_{ee}(E,\theta)$ [see Fig.\ref{fig2}] describe Andreev and
normal tunneling of an electron incident on the NS boundary with
the angle $\theta$ and the energy $E$ correspondingly into a hole
and an electron in the lead 2. Following the Landauer-B\"{u}ttiker
approach \cite{Takane_Ebisava,Anatram-Datta,Blanter-Buttiker}:
\begin{multline}
\label{I} I_2(V_1,V_2=0)= \frac{e}{\hbar}\int dE\sum_{_{\rm
channels}} (T_{ee}-T_{he})\times
\\
\times(f^{(1)}-f^{(2)}),
\end{multline}
where the sum is taken over channels (spin degrees of freedom are
included into channel definition); $f^{(1,2)}$ are distribution
functions in the leads 1,2; e.g., $f^{(2)}=n_F(E)=1/[1+\exp(\beta
E)]$, $f^{(1)}$ is not necessary a Fermi-function. We calculate
the transmission and reflection probabilities using Boguliubov
equations (BdG). The layers I are approximated by
$\delta$-barriers.
\begin{figure}[tb]
\begin{center}
\includegraphics[width=70mm]{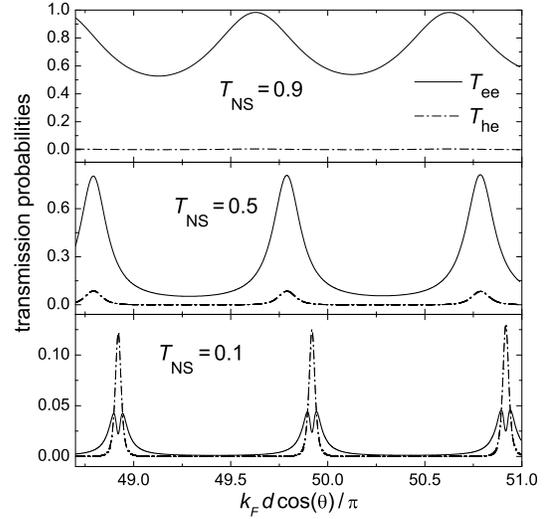}
\caption{Fig3. Resonances of Andreev and normal transmission
probabilities $T_{ee}(\theta)$, $T_{he}(\theta)$ of a NISIN
junction Fig.\ref{fig1}a at different transparencies
$T_{NS}(\theta)$ of the layers I. Parameters: $d/\xi=0.1$,
$E_{F}/\Delta=1000$, the energy $E=0$. Resonances correspond to
$k_F d\cos(\theta)/\pi=n$, $n=1,2,\ldots$. If $0<E<\Delta$ then
the shape of the graphs slightly change: the resonance peaks
become slightly asymmetric [same applies to the case when $T_{NS}$
are different but small]. It can be checked (even analytically)
that the areas under corresponding resonance peaks of $T_{ee}$ and
$T_{he}$ become equal at small $T_{NS}$.}\label{fig3}
\end{center}
\end{figure}
Quasiparticle motion parallel and perpendicular to the NS
interfaces can be decoupled \cite{Andreev,Radovic}. Matching
appropriate wave functions in the normal region and the
supercondctor we get $8\times 8$ linear system of equations for
the transmission amplitudes. Analytical progress can be made. It
follows that if there is no barrier at NS boundaries (except
$\Delta$) $T_{he}/T_{ee}\lesssim (\Delta/E_F)^2$ for any thickness
$d$ of the superconducting layer. This result is intuitively quite
clear because $\Delta\ll E_F$ can hardly reverse the direction of
the quasiparticle momentum being about $k_F$ \cite{Andreev,Imry}.
However if there are barriers at NS boundaries in addition to
$\Delta$ (e.g., insulating layers I) then the situation changes:
at certain $\theta$ the transmission probabilities have resonances
where $T_{he}\sim T_{ee}$. When the transmission probabilities of
the layers I, $T_{NS}^{(1,2)}\ll 1$, the areas under the resonance
peaks of $T_{eh}(\theta)$ and $T_{ee}(\theta)$ are nearly same and
\begin{gather}
\label{T_ee_eh} \langle T_{he}\rangle\thickapprox \langle
T_{ee}\rangle,
\end{gather}
where $\langle \ldots\rangle=\sum_{_{\rm
channels}}(\ldots)/N_{_{\rm channels}}\approx\int_0^{1}
(\ldots)d\cos^2\theta$. Eq. \eqref{T_ee_eh} is exact in first
nonvanishing order over $T_{NS}$. The resonances appear at $k_Fd
\cos(\theta_n)=\pi n$, $n=1,2\ldots$, give the leading
contributions to $\langle T_{he}\rangle$, $\langle T_{ee}\rangle$
and are responsible for Eq.\eqref{T_ee_eh}. The resonance width
$\Gamma\sim\min\{1,T_{NS},d/\xi\}$. Typical dependencies of
$T_{eh}(\theta)$ and $T_{ee}(\theta)$ from $\theta$ and $T_{NS}$
are illustrated in Fig.\ref{fig2}.

In fact $\theta$ is discrete variable; its particular value is
determined by the channel of the incident particle.  Equation
\eqref{T_ee_eh} is applicable when 1) $T_{NS}(\theta)$ slightly
change when $\theta$ changes from one channel to an adjacent one
and 2) change of $\theta$ from one channel to another should be
smaller than the resonance width. The condition 1) is fulfilled
typically when $T_{NS}(\theta)\ll 1$, the condition 2) requires
$\lambda_F/\sqrt{A}\ll \min\{1,T_{NS}(0),d/\xi\}$, where $A$ is
the junction surface area.

\begin{figure}[tb]
\begin{center}
\includegraphics[width=60mm]{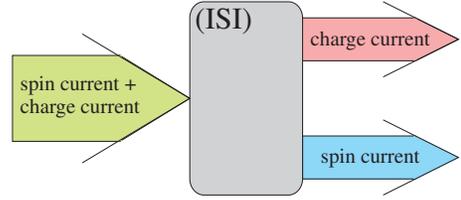}
\caption{Fig4. Isolator -- superconductor -- isolator- normal
metal structure can help to separate in space spin and charge
components of the current.} \label{fig4}
\end{center}
\end{figure}

It follows from Eqs.~(\ref{I}-\ref{T_ee_eh}) that subgap charge
injection from the lead 1 into the lead 2 in weak coupling regime
($T_{NS}\ll 1$) is suppressed: $I_2(V_1,V_2=0)=0$, because charge
currents of transmitted hole and electron quasiparticles
compensate each other in the lead 2; all the electron current
converts into Cooper-pair supercurrent in S. However if
spin-polarized current is injected from the lead 1 finite spin
current appears in the lead 2; transmitted electron and hole
quasiparticles contribute the spin current. XISIN structure with
$T_{NS}^{(1)}\ll T_{NS}^{(2)}\ll 1$ (this condition allows to
neglect the contribution to the charge current going in S from
Andreev reflection at N$_{1}$S surface), where X is current
``injector'', can play the role of the filter of spin and charge
currents, see fig.~\ref{fig4}. Equation for the spin current
follows from Eq. \eqref{I}:
\begin{multline}
\label{Is} I_2^{(s)}(V_1,V_2=0) = \frac{e}{2\hbar}\int
dE\sum_{_{\rm channels}}\sigma_1 (T_{ee}+T_{he})\times
\\
\times (f^{(1)}-f^{(2)}),
\end{multline}
where $\sigma_1=\pm 1$ labels spin degrees of freedom in X.
General feature of transmission probabilities $T$ and the current
--- their exponential suppression with $d/\xi$ when $d\gg\xi$ ($\xi$ is the
superconductor coherence length).

We show below that all the results discussed above remain true in
general NISIN structure with more complicated shape than in
Fig.\ref{fig1}a, (e.g. like in Fig.~\ref{fig1}b) no matter dirty
or clean.

In general a system of weakly coupled normal (ferromagnetic) and
superconducting layers can be described by the Hamiltonian:
$\hat{H}=\hat{H}_{1}+\hat{H}_{2}+\hat{H}_S+\hat{H}_T$, where
$\hat{H}_{1,2}$ refer to the electrodes N$_1$ and N$_2$, and
$\hat{H}_S$ to the superconductor. The tunnel Hamiltonian
$\hat{H}_T$, which we consider as a perturbation, is given by two
terms $\hat{H}_T=\hat{H}_{T}^{(1)}+\hat{H}_{T}^{(2)}$
corresponding to one-particle tunneling through each tunnel
junction:
\begin{gather}
\label{H} \hat{H}_{T}^{(i)}=\sum_{k,p}
\left\{\hat{a}^{(i)\dagger}_{ k}t_{ kp}^{(i)} \hat b_{ p}+ h. c.
\right\},
\end{gather} where indices
$i=1,2$ refer to normal (ferro) electrodes; $t_{k p}^{(i)}$ is the
matrix element for tunneling from the state $k=({\bf k},\sigma)$
in normal lead N$_i$ to the state $p=({\bf p},\sigma')$ in the
superconductor. The operators $\hat{a}^{(i)}_{k}$ and $\hat b_{p}$
correspond to quasiparticles in the leads and in the
superconductor, respectively.

\begin{figure}
\begin{center}
\includegraphics[width=60mm]{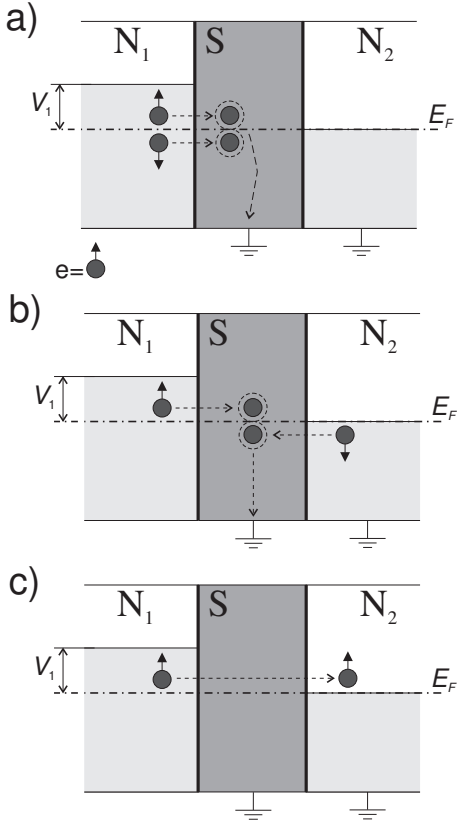}
\caption{Fig5. a) Direct Andreev tunneling (Andreev reflection),
b) Crossed Andreev tunneling (Andreev transmission), and c)
Elastic Cotunneling (normal transmission).} \label{fig5}
\end{center}
\end{figure}

The current can be expressed through the quasiparticle scattering
probabilities within the Landauer-B\"{u}ttiker approach. It is
possible to calculate the scattering probabilities within the
tight-binding model \eqref{H}, but it is more convenient to
describe the current on the language of electrons only: Andreev
transmission probability $T_{he}$ \eqref{I} is closely related to
the crossed Andreev (CA) tunneling rate
$\Gamma_{CA}^{S\leftarrow}(V_1,V_2)$ which shows how many electron
pairs tunnel per second from the leads 1 and 2 into the condensate
of the superconductor (each lead gives one electron into a pair)
and vice versa correspondingly, see Fig.~\ref{fig5}b, and
\cite{Falci}. Elastic co-tunneling rate
$\Gamma_{(EC)}^{2\leftarrow 1}$ corresponds to $T_{ee}$. Direct
Andreev tunneling rates, $\Gamma^{S\to 1(2)}_{(DA)}$ and
$\Gamma^{S\leftarrow 1(2)}_{(DA)}$ describe Andreev reflection in
the leads 1 and 2 [see, e.g., Fig.~\ref{fig5}a]. The current in
the lead 2 consists of two contributions: one, $I_2^{(i)}$, comes
from the electron injection from the lead 1 due to crossed-Andreev
and cotunneling processes, the other, $I_{2}^{(I)}$, -- from the
direct electron-tunneling between the lead and the superconductor.
Same applies for the lead 1.  Thus
$I_2(V_1,V_2)=I_2^{(I)}(V_1,V_2)+I_{2}^{(D)}(V_2)$, where
\begin{subequations}
   \begin{eqnarray}
\label{3} I_2^{(I)}(V_1,V_2)&=&\Gamma_{(EC)}-\Gamma_{(CA)},
\\
\Gamma^{(EC)}&=&\Gamma_{(EC)}^{1\leftarrow
2}-\Gamma_{(EC)}^{2\leftarrow 1},
\\
\Gamma^{(CA)}&=&\Gamma_{(CA)}^{S\to
1,2}-\Gamma_{(CA)}^{S\leftarrow 1,2},
\\
I_2^{(D)}(V_2)&=&\Gamma_{(DA)}^{S\leftarrow 2}-\Gamma^{S\to
2}_{(DA)}.
   \end{eqnarray}
\end{subequations}
Using the Fermi Golden rule, the rates can be found in  second
order in the tunneling amplitude $t_{k,p}$. Following the approach
described in Ref.~\cite{Falci,HekkingNazarov,Burmistrov}, we
finally obtain
\begin{multline}
\label{Gamma} \Gamma_{CA}^{S\leftarrow 12}(V_1,V_2)=4\pi^3 \int
d\xi \sum_\sigma
n_{\sigma}^{(1)}(\xi-V_1)n_{-\sigma}^{(2)}(-\xi-V_2)
\\\times
\frac {\Delta^2}{[\Delta^2-\xi^2]}\tilde
\Xi^{(CA)}_\sigma(2\sqrt{\Delta^2-\xi^2}),
\end{multline}
where $n^{(i)}$ is the distribution function in the lead $i=1,2$.
Hereafter we take $\hbar=1,e=1$ [we do not assume $n^{(i)}$ to be
only equilibrium Fermi function]. The rate
$\Gamma_{CA}^{S\rightarrow 12}$ can be obtained from the
expression for $\Gamma_{CA}^{S\leftarrow 12}$ by substitution of
$(1-n)$ for $n$.

The kernel $\tilde \Xi^{(CA)}_\sigma(s)\equiv \int_0^\infty dt
\Xi^{(CA)}_\sigma(t)e^{-st}$ is the Laplace transform of
$\Xi^{(CA)}_\sigma(t)$. It can be expressed through the classical
probability $P(X_{1},\hat p_1; X_{2},\hat p_2,t)$ meaning that an
electron with the momentum directed along $\hat p_1$ initially
located at the point $X_1$ near the NS boundary arrives at time
$t$ at some point $X_2$ near the NS boundary with the momentum
directed along $\hat p_2$ spreading in the superconducting region
as follows
\begin{multline*}
 \Xi_\sigma^{(CA)}(t)=
\\
=\frac{1}{8 \pi^3 e^4 \nu_S}\int d\hat p_{1,2}\int dX_{1,2}
P(X_{1},\hat p_1; X_{2},\hat p_2,t)\times
\\
\left\{G^{(1)}(X_1,\hat p_1,\sigma)G^{(2)}(X_2,\hat p_2,
\sigma)\sin^2\left(\frac{\theta(X_1,X_2)} 2\right)+\right.
\\
+\left.G^{(1)}(X_1,\hat p_1,\sigma)G^{(2)}(X_2,\hat p_2,
-\sigma)\cos^2\left(\frac{\theta(X_1,X_2)} 2\right)\right\}.
\end{multline*}
Here the spatial integration is performed over the N$_1$S and
N$_2$S surfaces. We choose the spin quantization axis in the
direction of the local magnetization in the terminal N$_{1(2)}$.
The quasiclassical probabilities $G^{(i)}(X,\hat p,\sigma)$,
$i=1,2$ for the electron with spin polarization $\sigma$ tunneling
from the terminal N$_i$ to the superconductor are normalized in
such way that the junction normal conductance per unit area
$g_{\sigma}^{(i)}(X)$ and the total normal conductance
$G_{N}^{(i)}$ are determined
as~\cite{HekkingNazarov,AverinNazarov}
\begin{gather*}
g_{\sigma}^{(i)}(X) = \int d\hat p\, G^{(i)}(X,\hat
p,\sigma),\quad G_{N}^{(i)} = \int d X \sum \limits_{\sigma}
g_{\sigma}^{(i)}(X).
\end{gather*}
Then the normal conductance per unit area, discussed above, is
defined as $g_N^{(i)}=G_N^{(i)}/{\cal A}$, where ${\cal A}$ is the
surface area of the junction. Symbol $\theta(X_1,X_2)$ is the
angle between the magnetizations of the terminals N$_1$ and N$_2$
at points $X_1$ and $X_2$ near the junction surface. If electrons
in N$_1$ and N$_2$ are not polarized then $\theta=0$.

In a similar way:
\begin{multline*}
\Gamma_{(EC)}^{1\rightarrow 2}=4\pi^3 \int d\xi \sum_\sigma
n_{\sigma}^{(1)}(\xi-V_1)(1-n_{\sigma}^{(2)}(\xi-V_2))
\\\times
\frac {\Delta^2}{[\Delta^2-\xi^2]}\tilde
\Xi^{(EC)}_\sigma(2\sqrt{\Delta^2-\xi^2}),
\end{multline*}
where
\begin{multline}
\label{Xi_EC} \Xi_\sigma^{(EC)}(t)=
\\
=\frac{1}{8 \pi^3 e^4 \nu_S}\int d\hat p_{1,2}\int dX_{1,2}
P(X_{1},\hat p_1; X_{2},\hat p_2,t)\times
\\
\left\{G^{(1)}(X_1,\hat p_1,\sigma)G^{(2)}(X_2,\hat p_2,
-\sigma)\sin^2\left(\frac{\theta(X_1,X_2)} 2\right)+\right.
\\
+\left.G^{(1)}(X_1,\hat p_1,\sigma)G^{(2)}(X_2,\hat p_2,
\sigma)\cos^2\left(\frac{\theta(X_1,X_2)} 2\right)\right\}.
\end{multline}
The rate $\Gamma_{(EC)}^{1\leftarrow 2}$ can be obtained from the
expression for $\Gamma_{(EC)}^{1\rightarrow 2}$ by substitution of
$(1-n)$ for $n$. DA-rates are written in \cite{HekkingNazarov}.
Equations (\ref{3}-\ref{Xi_EC}) derived here allow to describe
transport properties of many types of junctions.

Consider FISIN junction with biased ferromagnet with the respect
to the superconductor, the normal metal N has same voltage as S.
So the ferromagnet plays the role of a current ``injector'';
electrons coming from F are distributed with some distribution
function $n^{(1)}$. Electrons in the deep of the terminal N are
Fermi-distributed. It follows from Eqs.(\ref{3}-\ref{Xi_EC}) that
contributions to the current from EC and CA processes compensate
each other for subgap voltages so $I_N(V_F\neq 0,0)=0$. However
spin current is finite:
\begin{multline}
I^{(\textrm{spin})}_N(V_F,0)=4\pi^3 \int d\xi \sum_\sigma
\sigma[n_{\sigma}^{(1)}(\xi-V_F)-n^{(2)}(\xi)]
\\\times
\frac {\Delta^2}{[\Delta^2-\xi^2]} \frac{1}{8 \pi^3 e^4 \nu_S}
\int d\hat p_{1,2}\int dX_{1,2} \times
\\
P(X_{1},\hat p_1; X_{2},\hat p_2,t) G^{(1)}(X_1,\hat
p_1,\sigma)G^{(2)}(X_2,\hat p_2).
\end{multline}

Finally we consider a FISIF junction. It was shown in \cite{Falci}
that in this junction $I_2(V_1,0)\neq 0$ and $I_2(V_1,0)$ changes
its sign when the ferromagnetic terminals change their orientation
from parallel to antiparallel. Naively it can be supposed that in
a FISIN junction where F is a current injector, S, N are grounded
spin accumulation at the interface of the normal metal would lead
to spin-splitting of the density of states in N and a charge
current. However this is not so, this corrections are of higher
order over tunneling amplitudes than the processes in
Fig.~\ref{fig5} and can  be neglected because we assume that
tunneling amplitudes are small.

It was also noted in \cite{Falci,Melin} that the cross conductance
$G_{12}\equiv\partial_{V_1} I_2(V_1,0)|_{V_1=0}$ is suppressed in
a FISIF structure as $1/(k_F r)^2$ when the characteristic
distance between the ferromagnets $r<\xi$. In dirty regime there
is no conductance suppression at atomic-scales. Consider, for
instance, the layout sketched in Fig.~\ref{fig1}b; the width $d$
of the superconducting film is supposed to be smaller than $\xi$.
According to Eqs.(\ref{3}-\ref{Xi_EC}) the cross-conductance
dependence from the distance $r$ is determined by the Laplace
transform $\tilde P(s=2\sqrt{\Delta})$ of the probability
$P(r,t)=\exp(-r^2/4D|t|)/4\pi d |t|$, where $D$ is diffusion
constant in the superconductor, $d<\xi$. When $\lambda_F\ll
r<\xi$, $G_{12}\sim\tilde P\sim \ln(r/\xi)$ and if $r\gg \xi$,
$G_{12}\sim\tilde P\sim \exp(-r/\xi)$. When the superconductor is
bulk ($d>\xi$) similarly we find $G_{12}\sim \xi/r$, $\lambda_F\ll
r<\xi$. All considerations above apply also for CA- and EC-rates.
Thus it is practically more convenient to measure finite effects
related to electron subgap tunneling through a superconductor when
it is dirty rather than clean. In dirty case the terminals are not
restricted to be as close as $\lambda_F$ like in clean case but
closer then $\xi\gg\lambda_F$.

We are grateful to M. Mar'enko, Yu.V. Nazarov, V.V. Ryazanov, M.V.
Feigelman, A.S. Iosselevich, and Ya.V. Fominov for stimulating
discussions. M. Mar'enko paid my attention to suppression of the
zero bias cross conductance in  dirty NISIN junctions (of certain
type) and the long-range decay of the EC and CA-rates with the
characteristic distance between the normal terminals which
appeared important for reviewing in general case spin and charge
transport in the superconducting junctions with weak coupling to
the normal (ferromagnetic) terminals. After the paper was nearly
completed I got the information that spin injection in a normal
layer of a FISIN junction was mention in one sentence of
Ref.\cite{Moriond}. We thank to D. Feinberg for criticism and
pointing our attention to Ref.\cite{Moriond}.
We wish to thank RFBR (project No.  03-02-16677), the Swiss NSF
and Russian Ministry of Science.


\begin{thebibliography}{99}
\bibitem{Deutscher} G.~Deutscher, D.~Feinberg, Appl. Phys. Lett. \textbf{76}, 487
(2000).

\bibitem{Falci} G.~Falci, D.~Feinberg, F.\,W.\,J.~Hekking, Europhys. Lett.
\textbf{54}, 225 (2001).

\bibitem{Melin} R. Melin, D. Feinberg,
Eur. Phys. J. B 26, 101 (2002).

\bibitem{Lesovik} G.B.\ Lesovik, T.\ Martin, and G.\ Blatter,
Eur. Phys. J. B {\bf 24}, 287 (2001); P. Recher, E. V. Sukhorukov,
and D. Loss, Phys. Rev. B {\bf 63}, 165314 (2001); N.M.
Chtchelkatchev, G. Blatter, G.B. Lesovik \textit{et al}., Phys.
Rev. B \textbf{66}, R161320 (2002); M.\,S.~Choi, C.~Bruder, and
D.~Loss, Phys. Rev. B \textbf{62}, 13569 (2000).

\bibitem{Jedema} F. J. Jedema, B. J. van Wees, B. H. Hoving \textit{et al.},
Phys. Rev. B \textbf{60}, 16549 (1999).

\bibitem{q_information}D. Bouwmeester, A. Ekert, and A. Zeilinger, {\it The Physics of Quantum Information: Quantum Cryptography, Quantum
Teleportation, Quantum Computations} (Springer-Verlag, Berlin,
2000).

\bibitem{Fazio} F. Giazotto, F. Taddei, R. Fazio \textit{et  al}.,
Appl. Phys. Lett. \textbf{82}, 2449 (2003).

\bibitem{Takane_Ebisava} C.J.Lambert, J.Phys.C {\bf 3}, 6579(1991); Y.Takane and H.Ebisawa, J.Phys.Soc.Jpn.
{\bf 61}, 1685 (1992).

\bibitem{Anatram-Datta} M.P.\ Anantram and S.\ Datta,
Phys.\ Rev.\ B {\bf 53}, 16390~(1996).

\bibitem{Blanter-Buttiker} Ya. Blanter, M. Buttiker, Phys. Rep. {\bf 336}, 1
(2000).

\bibitem{Andreev} A.\,F.~Andreev, Sov. Phys. JETP \textbf{19}, 1228 (1964).
%

\bibitem{Radovic} M. Bo\v{z}ovi\'{c} and Z. Radovi\'{c}, Phys.
Rev. B \textbf{66}, 134524 (2002).

\bibitem{Imry}
Y. Imry, Introduction to mesoscopic physics, Oxford University
Press, 1997.

\bibitem{HekkingNazarov} F.\,W.~Hekking, Yu.\,V.~Nazarov, Phys.
Rev. B \textbf{49}, 6847 (1994).

\bibitem{AverinNazarov} D.V. Averin and Yu.V. Nazarov,
Phys. Rev. Lett. {\bf 65}, 2446 (1990).

\bibitem{Burmistrov} N.M. Chtchelkatchev \textit{et al},
cond-mat/0303014.

\bibitem{Moriond} D. Feinberg, G. Deutscher, G. Falci \textit{et al}., proceedings of
Rencontres de Moriond 2001, p. 535, eds T. Martin, G. Montambaux
and J. Tran Thanh Van (EDPSciences 2001).


\end{thebibliography}
\end{document}